\documentclass[10pt]{revtex4}
\usepackage{graphicx}

\begin{document}

\title{Observing The Observer II: Can I know I am in a superposition and still be in a superposition?}
\author{Vlatko Vedral}
\affiliation{Clarendon Laboratory, University of Oxford, Parks Road, Oxford OX1 3PU, United Kingdom\\Centre for Quantum Technologies, National University of Singapore, 3 Science Drive 2, Singapore 117543\\
Department of Physics, National University of Singapore, 2 Science Drive 3, Singapore 117542}

\begin{abstract}
The answer will be a ``yes" (despite looking like a violation of the Uncertainty Principle). 
\end{abstract}

\maketitle

In a previous exposition I argued that the idea that quantum mechanics applies to everything in the universe, even to us humans, can lead to some interesting conclusions \cite{Vedral-partI,Vedral}. I will repeat part of the argument already presented in \cite{Vedral-partI} simply for the sake of completeness
and then extend it to argue that we can be in a superposition while knowing that we are.

Consider Deutsch's variant \cite{Deutsch} of the Schr\"odinger cat thought experiment that builds on Wigner's ideas \cite{Wigner}. Suppose that a very able experimental physicist, Alice, puts her friend Bob inside a room with a cat, a radioactive atom and cat poison that gets released if the atom decays. The point of having a human there is that we can communicate with him. As far as Alice is concerned, the atom enters into a state of being both decayed and not decayed, so that the cat is both dead and alive (that's where Schr\"odinger stops). 

Bob, however, can directly observe the cat and sees it as one or the other. This is something we know from everyday experience: we never see dead and alive cats. To confirm this, Alice slips a piece of paper under the door asking Bob whether the cat is in a definite state. He answers, ``Yes, I see a definite state of the cat".

At this point, mathematically speaking, the state of the system has changed from the initial state
\begin{equation}
|\Psi_i\rangle = |\textnormal{no-decay}> |\textnormal{poison in the bottle}> |\textnormal{cat alive}> |\textnormal{Bob sees alive cat}> |\textnormal{blank piece of paper}>
\end{equation}
to the state (from Alice's global perspective):
\begin{eqnarray}
|\Psi_{1/2}\rangle = & & (|\textnormal{decay}> |\textnormal{poison released}> |\textnormal{cat dead}> |\textnormal{Bob sees dead cat}> +\nonumber \\
& & |\textnormal{no-decay}> |\textnormal{poison still in the bottle}> |\textnormal{cat alive}> |\textnormal{Bob sees alive cat}>) \otimes \nonumber \\
& & |\textnormal{paper says: ”yes, I see a definite state of the cat”}> 
\label{cat}
\end{eqnarray}
I am assuming that, because Alice's laboratory is isolated, every transformation leading up to this state is unitary. This includes the decay, the poison release, the killing of the cat and Bob's observation - Alice has a perfect quantum coherent control of the experiment. 

Note that Alice does not ask whether the cat is dead or alive because for her that would force the outcome or, as some physicists might say, ``collapse” the state (this is exactly what happens in Wigner's version, where he communicates the state to a friend, who communicates to another friend and so on...). She is content observing that Bob sees the cat either alive or dead and does not ask which it is. Because Alice avoided collapsing the state (in other words, she did not get entangled to her experiment), quantum theory holds that slipping the paper under the door was a reversible act. She can undo all the steps she took since each of them is just a unitary transformation. In other words, the paper itself also does not get entangled to the rest of the laboratory. 

When Alice reverses the evolution, if the cat was dead, it would now be alive, the poison would be in the bottle, the particle would not have decayed and Bob would have no memory of ever seeing a dead cat. If the cat was alive, it would also come back to the same state (everything, in other words, comes back to the starting state where the atom has not decayed, the poison is in the bottle, the cat is alive and Bob sees alive cat and has no memory of the experiment he was subjected to). 

And yet one trace remains: the piece of paper saying ``yes, I see a definite state of the cat". Alice can undo Bob's observation in a way that does not also undo the writing on the paper. The paper remains as proof that Bob had observed the cat as definitely alive or dead half way through the experiment. (Note that I remain interpretation neutral. A Many Worlds supporter would say that there are two copies of Bob, one that observes a dead cat and one that sees alive cat; a Copenhagen or Quantum Bayesian supporter could say that relative to one state of Bob the cat is dead, while, relative to the other, it is alive - either way, supporters of any interpretation ought to make the same predictions in this experiment).

This is where I stopped last time \cite{Vedral-partI}. 

However, before reversing the evolution (and after Alice receives a communication from Bob that he sees a definitive outcome), Alice can actually communicate with Bob again (the first communication being Alice's question whether Bob sees a definite state). This time she says to him (by slipping another piece of paper under the door): ``From your reply I know you see a definite outcome, but I am now telling you that you are nevertheless in a superposed state of seeing both outcomes. Or, more precisely, there is a version of you (or of your consciousness or whatever) that sees the cat dead and one that sees the cat alive". (something similar was discussed by Albert in \cite{Albert}). Even better, if Bob is himself a quantum physicist, Alice could just write down the equation describing the state of the laboratory on the same piece of paper. This equation would just be the same as Eq. (\ref{cat}). 

Bob, if he trusts Alice (and why shouldn't he? - she is both a good friend and a good physicist), might be shocked. He might think ``I see a definitive outcome, so how can I still be in a superposition?". This sounds like a double slit experiment in which each particle goes through only one slit at a time and yet we obtain an interference pattern at the end. This would be a clear violation of the Uncertainty Principle. 

The answer to this apparent conundrum is, of course, that Bob is not in a superposition. Rather, he is entangled to the cat and the poison and the decayed atom, exactly as above. And, being maximally entangled to something means not being in a superposition but in a mixed state. So Bob now knows he exists in two different ``worlds" (or rather, each version knows about the other), yet each of the two versions of him feels as though they are safely operating within one world only. Note that even though this language sounds "manyworldish" what we are discussing is simply an experimental question. All interpretations of quantum physics will have to agree on the outcome albeit they might be using different jargon to describe the situation. 

In fact, Alice can perform measurements to confirm that Bob is in the entangled state in Eq. (\ref{cat}) without collapsing the state and then send the experimental results to Bob to dispell any doubts (of course, Bob would have to trust her that she performed the relevant experiments and that the results he has received from her are indeed genuine). 

But, let's push this a bit further. Suppose Alice, after telling Bob he exists in two worlds, offers him a bet. She tells him that she will now perform a reversal of the experiment and thereby interfere the two possibilities. She says to him: ``What will you see at the end when I am done with the reversal?". We seem to be facing two logical possibilities. One is that Bob will be in the same state as at the beginning of the whole experiment (and the cat will be alive, the poison in the bottle and the atom not-decayed). This is what we argued quantum physics would lead to. The other is that he will see the cat dead, the poison out and the atom decayed. But how can this second possibility happen?

It cannot, unless there is a collapse of quantum physics due to Bob's observation or due to something else. Note, however, that we are assuming that Alice's experimental skills are faultless. But, if Bob really thinks that, despite Alice's communication informing him he is in two states at once, he feels that this cannot be possible and goes along with his feeling that he is definitely in one of the two worlds - this is why he sees a definite outcome after all - then he could, in principle, end up after her reversal in the dead-cat branch. Bob ending up in it would be a proof that quantum physics cannot describe the overall experiment (in which case all interpretations of quantum physics fail and not just the Many Worlds). 

So Bob, if he believes Alice and that quantum physics will uphold, should bet on seeing the cat alive at the very end. If quantum physics is correct, Bob cannot interfere his two states half way through the experiment to check. Neither of his two versions has ``access" to the other one. This is presumably why each of his versions feels like being in a definitive world.  Alice can do so because she is in charge of the experiment and can remain disentangled from the subjects of the experiment. Of course, this is a difficult experiment to perform, unless somewhere within our brain there is a ``quantum bit worth of perception" that, at least sometimes, is capable of encoding our observations and can be fully quantum mechanical (for a sufficiently long time to perform the experiment).  

How would this work? I have no idea, but imagine something like the following (I am now speculating on the verge of being irresponsible).  There is a famous two-dimensional line drawing of a cube first brought to light by Swiss psychologist Louis Necker. Even though there are many logical interpretations of this picture (many actually impossible in 3 dimensions) the brain does not see any ambiguity in this two dimensional drawing. In fact, we see only two of the possible images (with one of the surfaces at the front or at the back). Most people see one of these first (this is apparently to do with the way we understand perspective) and then, after a few seconds or so, revert to the other one, thereafter continuing 
to flip back and forth between the two. 

Suppose now that the two images of the Necker cube are stored in our brain as two distinct quantum states. Admittedly, these states could be very complex, in the sense of involving many atoms and interactions between them. The mind then switches between the two physical states, which is like a logical operation of a flip from zero to one and back to zero. The question is: could it be that the switching process is actually quantum mechanical and that during the transition our brain is actually storing a superposition of the two images? (I discuss this at greater length in \cite{Vedral-micro-macro}).

If so, this could maybe give us a small window of opportunity to be able to do something like Alice. We could perhaps confirm that Bob sees a definitive version of the Necker cube and then possibly undo this observation (providing we understand enough about how it is stored) thereby demonstrating that Bob has literally been, quantumly speaking, in two minds at the same time. This would not mean that quantum physics was necessary for perception, even less for consciousness, however, it would, as far as I am concerned, constitute the most remarkable example of a macroscopic quantum effect to date.

\textit{Acknowledgments}: The author acknowledges funding from the National Research Foundation
(Singapore), the Ministry of Education (Singapore), the Engineering and Physical Sciences Research Council (UK), the Templeton Foundation, the Leverhulme Trust, the Oxford Martin School, and Wolfson
College, University of Oxford. This research is also supported by the National Research Foundation, Prime Minister’s Office, 
Singapore under its Competitive Research Programme (CRP Award No. NRF- CRP14-2014-02) and administered by 
Centre for Quantum Technologies, National University of Singapore.

\end{document}